 
\input harvmac.tex
\overfullrule=0pt
\def\p{\partial}

\def\p{\partial}
\def\unlockat{\catcode`\@=11}
\def\lockat{\catcode`\@=12}
\unlockat
\def\newsec#1{\global\advance\secno by1\message{(\the\secno. #1)}
\global\subsecno=0\global\subsubsecno=0\eqnres@t\noindent
{\bf\the\secno. #1}
\writetoca{{\secsym} {#1}}\par\nobreak\medskip\nobreak}
\global\newcount\subsecno \global\subsecno=0
\def\subsec#1{\global\advance\subsecno
by1\message{(\secsym\the\subsecno. #1)}
\ifnum\lastpenalty>9000\else\bigbreak\fi\global\subsubsecno=0
\noindent{\it\secsym\the\subsecno. #1}
\writetoca{\string\quad {\secsym\the\subsecno.} {#1}}
\par\nobreak\medskip\nobreak}
\global\newcount\subsubsecno \global\subsubsecno=0
\def\subsubsec#1{\global\advance\subsubsecno by1
\message{(\secsym\the\subsecno.\the\subsubsecno. #1)}
\ifnum\lastpenalty>9000\else\bigbreak\fi
\noindent\quad{\secsym\the\subsecno.\the\subsubsecno.}{#1}
\writetoca{\string\qquad{\secsym\the\subsecno.\the\subsubsecno.}{#1}}
\par\nobreak\medskip\nobreak}

\def\subsubseclab#1{\DefWarn#1\xdef
#1{\noexpand\hyperref{}{subsubsection}%
{\secsym\the\subsecno.\the\subsubsecno}%
{\secsym\the\subsecno.\the\subsubsecno}}%
\writedef{#1\leftbracket#1}\wrlabeL{#1=#1}}
\lockat


\def\IL{\relax{\rm I\kern-.18em L}}
\def\IH{\relax{\rm I\kern-.18em H}}
\def\IR{\relax{\rm I\kern-.18em R}}
\def\IC{\relax\hbox{$\inbar\kern-.3em{\rm C}$}}
\def\IZ{\relax\ifmmode\mathchoice
{\hbox{\cmss Z\kern-.4em Z}}{\hbox{\cmss Z\kern-.4em Z}}
{\lower.9pt\hbox{\cmsss Z\kern-.4em Z}}
{\lower1.2pt\hbox{\cmsss Z\kern-.4em Z}}\else{\cmss Z\kern-.4em
Z}\fi}

\def\CF {{\cal F}}

\def\CK {{\cal K}}

\def\CO {{\cal O}}

\def\CH {{\cal H}}
\def\CC {{\cal C}}
\def\CB {{\cal B}}
\def\CS {{\cal S}}
\def\CA{{\cal A}}


\def\CO {{\cal O}}

\def\CS {{\cal S }}
\def\ch{{\rm ch}}

\font\manual=manfnt \def\dbend{\lower3.5pt\hbox{\manual\char127}}

\def\IZ{\relax\ifmmode\mathchoice
{\hbox{\cmss Z\kern-.4em Z}}{\hbox{\cmss Z\kern-.4em Z}}
{\lower.9pt\hbox{\cmsss Z\kern-.4em Z}}
{\lower1.2pt\hbox{\cmsss Z\kern-.4em Z}}\else{\cmss Z\kern-.4em
Z}\fi}
\def\half {{1\over 2}}

\def\p{\partial}

\def\CO {{\cal O}}

\def\CS {{\cal S }}
\def\ch{{\rm ch}}


\def\IZ{\relax\ifmmode\mathchoice
{\hbox{\cmss Z\kern-.4em Z}}{\hbox{\cmss Z\kern-.4em Z}}
{\lower.9pt\hbox{\cmsss Z\kern-.4em Z}}
{\lower1.2pt\hbox{\cmsss Z\kern-.4em Z}}\else{\cmss Z\kern-.4em
Z}\fi}
\def\IA{\relax{\rm I\kern-.18em A}}
\def\IB{\relax{\rm I\kern-.18em B}}
\def\IC{{\relax\hbox{$\inbar\kern-.3em{\rm C}$}}}
\def\ID{\relax{\rm I\kern-.18em D}}
\def\IE{\relax{\rm I\kern-.18em E}}
\def\IF{\relax{\rm I\kern-.18em F}}
\def\IG{\relax\hbox{$\inbar\kern-.3em{\rm G}$}}
\def\IGa{\relax\hbox{${\rm I}\kern-.18em\Gamma$}}
\def\IH{\relax{\rm I\kern-.18em H}}
\def\II{\relax{\rm I\kern-.18em I}}
\def\IK{\relax{\rm I\kern-.18em K}}
\def\IP{\relax{\rm I\kern-.18em P}}

\def\inbar{\,\vrule height1.5ex width.4pt depth0pt}

\def\p{\partial}

\font\cmss=cmss10 \font\cmsss=cmss10 at 7pt
\def\IR{\relax{\rm I\kern-.18em R}}

\def\Tr{\rm Tr}


\def\boxit#1{\vbox{\hrule\hbox{\vrule\kern8pt
\vbox{\hbox{\kern8pt}\hbox{\vbox{#1}}\hbox{\kern8pt}}
\kern8pt\vrule}\hrule}}
\def\mathboxit#1{\vbox{\hrule\hbox{\vrule\kern8pt\vbox{\kern8pt
\hbox{$\displaystyle #1$}\kern8pt}\kern8pt\vrule}\hrule}}


\def\ap{\alpha'}

\def\inbar{\,\vrule height1.5ex width.4pt depth0pt}

\def\p{\partial}

\font\cmss=cmss10 \font\cmsss=cmss10 at 7pt
\def\IR{\relax{\rm I\kern-.18em R}}

\def\Tr{\rm Tr}


\lref\gms{R. Gopakumar, S. Minwalla, and A. Strominger, ``Noncommutative
Solitons,'' JHEP {\bf 0005}:020,2000; hep-th/0003160.}
\lref\sen{A. Sen, ``Non-BPS States and Branes in String Theory,''
hep-th/9904207 and references therein.}
\lref\wittenk{E. Witten, ``$D$-Branes And $K$-Theory,''
JHEP {\bf 9812}:019, 1998; hep-th/9810188.}
 \lref\dmr{K. Dasgupta, S. Mukhi and G. Rajesh, ``Noncommutative Tachyons,''
JHEP {\bf 0006}:022,2000; hep-th/0005006.}
\lref\hklm{J. Harvey, P. Kraus, F. Larsen, and E. Martinec,
``D-branes and Strings as Non-commutative Solitons,''
JHEP{\bf 0007}:042, 2000; hep-th/0005031.}
\lref\wittencomment{E. Witten, ``Noncommutative tachyons and string
field
theory,'' hep-th/0006071.}
\lref\hk{J. A. Harvey and P. Kraus, ``D-Branes as Lumps in Bosonic
Open String Field Theory,'' JHEP {\bf 0004}:012,2000; hep-th/0002117.}
\lref\MJMT{R. de Mello Koch, A. Jevicki, M. Mihailescu and R. Tatar,
``Lumps and P-Branes in Open String Field Theory,'' hep-th/0003031.}
\lref\sw{N. Seiberg and E. Witten, ``String Theory and Noncommutative
Geometry,'' JHEP {\bf 9909}:032,1999; hep-th/9908142.}
\lref\turaev{V. Turaev, ``Homotopy field theory in
dimension 2 and group-algebras,'' math.QA/9910010.}
\lref\dw{R. Dijkgraaf and E. Witten, ``Topological
Gauge Theories and Group Cohomology,'' CMP}
\lref\blackadar{B. Blackadar, {\it K-Theory for Operator
Algebras}, MSRI Publications 5, Cambridge Univ. Press,
1998.}
\lref\boutet{L. Boutet de Monvel, ``On the index of Toeplitz
operators of several complex variables,'' Inv. Math. {\bf 50}
(1979) 249.}
\lref\wittenstrings{E. Witten, ``Overview of K-theory applied to
strings,'' hep-th/0007175.}
\lref\asskewadjoint{M.F. Atiyah and I.M. Singer, ``Index
Theory for Skew-Adjoint Fredholm Operators,'' Inst. Hautes Études Sci. Publ.
Math.
No. 37, 1969 5--26.}
\lref\periwala{V. Periwal, ``D-brane charges and K-homology,''
hep-th/0006223.}
\lref\periwalb{V. Periwal, ``Deformation Quantization as the Origin
of D-Brane Nonabelian Degrees of Freedom,'' hep-th/0008046.}
\lref\lns{A. Losev, N. Nekrasov, and S. Shatashvili, ``The Freckled
Instantons,'' To appear in Yuri Golfand Memorial Volume; hep-th/9908204.}
\lref\baumdouglas{P. Baum and R.G. Douglas, ``K Homology and Index
Theory,'' Proc. Symp. Pure Math. {\bf 38}(1982) 117.}
\lref\bottandtu{R. Bott and L. Tu, {\it Differential forms in algebraic
topology}, Springer 1982.}
\lref\atiyahk{M. Atiyah, {\it K-theory}, Addison-Wesley, 1989. }
\lref\janich{Janich. Ref. in Segal's paper}
\lref\connes{A. Connes, {\it Noncommutative Geometry}, Academic
Press, 1994.}
\lref\horava{P. Horava, ``Type II D-Branes, K-Theory, and Matrix
Theory,'' Adv. Theor. Math. Phys. {\bf 2} (1999) 1373; hep-th/9812135.}
\lref\ks{V. A. Kostelecky and S. Samuel, ``On a Nonperturbative Vacuum
for the Open Bosonic String,'' Nucl. Phys. {\bf B336} (1990) 263.}
\lref\sz{A. Sen and B. Zwiebach, ``Tachyon Condensation in String Field
Theory,'' JHEP {\bf 0003}:002,2000; hep-th/9912249.}
\lref\bsz{N. Berkovits, A. Sen and B. Zwiebach, ``Tachyon Condensation
in Superstring Field Theory,'' hep-th/0002211.}
\lref\wt{N. Moeller and W. Taylor, ``Level truncation and the tachyon in open
bosonic string field theory,'' Nucl. Phys. {\bf B583}(2000)105;
hep-th/0002237.}
\lref\abs{Atiyah, M. F., Bott, R., Shapiro, A.  ``Clifford modules,''
Topology 3 1964 suppl. 1, 3--38. }
\lref\bdf{L.G.  Brown, R.G.  Douglas and P.A. Fillmore, ``Unitary equivalence
modulo the compact operators and extensions of $C^*$-algebras,''
Proc. of a Conference on Operator Theory, pp. 58-128, Lect. Note
in Math. Vol. 345, 1973; ``Extensions of $C^*$-algebras and K-homology,''
Ann. Math. {\bf 105}
(1977) 265.}
\lref\lru{U. Lindstrom, M. Rocek and R. von Unge, ``Non-commutative
Soliton Scattering,'' hep-th/0008108.}
\lref\wegge{N.E. Wegge-Olsen, {\it K-theory and $C^*$-algebras}, Oxford Univ.
Press, 1993.}
\lref\rosenberg{J. Rosenberg, ``Homological Invariants of Extensions of
$C^*$-algebras,'' Proc. Symp. Pure Math {\bf 38} (1982) 35.}
\lref\douglasnc{M. Douglas, ``D-Branes in Curved Space,'' hep-th/9703056;
``D-branes and Matrix Theory in Curved Space,''hep-th/9707228;
``Two Lectures on D-Geometry and Noncommutative Geometry,''
hep-th/9901146.}
\lref\bouwknegt{P. Bouwknegt and V. Mathai, ``D-Branes,
B-fields, and Twisted K-Theory,'' hep-th/0002023.}
\lref\atiyahsegal{M. Atiyah and G.B. Segal, unpublished.}
\lref\michigantalks{Talks by J. Harvey and G. Moore at Strings 2000,
http://feynman.physics.lsa.umich.edu/strings2000/schedule.html.}
\lref\segaltalks{See ITP lectures by G. Segal,
http://doug-pc.itp.ucsb.edu/online/geom99/. }
\lref\singertalks{
http://online.itp.ucsb.edu/online/geom/singer1/. }
\lref\rieffel{M.A. Rieffel, ``On the uniqueness of the Heisenberg
commutation relations,'' Duke Math. J. {\bf 39}(1972)745.}
\lref\janich{K. J\"anich, ``Vektorraumbundel und der Raum
der Fredholm=Operatoren,'' Math. Annalen {\bf 161}(1965)129.}
\lref\bfss{T. Banks, W. Fischler, S. H. Shenker and L. Susskind,
``M Theory as a Matrix Model: A Conjecture,'' Phys. Rev.
{\bf D55} (1997) 5112; hep-th/9610043.}
\lref\mm{R. Minasian and G. Moore,``K Theory and Ramond-Ramond Charge,''
JHEP {\bf 9711}:002, 1997; hep-th/9710230.}
\lref\myers{R. C. Myers, ``Dielectric Branes,'' JHEP {\bf 9912}:022,
1999; hep-th/9910053.}
\lref\ft{E. S. Fradkin and A.A. Tseytlin, ``Nonlinear Electrodynamics
>From Quantized Strings,'' Phys. Lett. {\bf 163B} (1985) 123.}
\lref\acny{A. Abouelsaood, C. G.Callan, C. R. Nappi and S. A. Yost,
``Open Strings in Background Gauge Fields,'' Nucl. Phys.
{\bf B280} (1987) 599.}
\lref\garousi{M. R. Garousi, ``Tachyon couplings on non-BPS D-branes and
Dirac-Born-Infeld action,'' Nucl.Phys. {\bf B584}
(2000) 284; hep-th/0003122.}
\lref\cds{A. Connes, M. R. Douglas and A. Schwarz, ``Noncommutative Geometry
and Matrix Theory:Compactification on Tori,'' JHEP {\bf 9802}:003 (1998);
hep-th/9711162.}
\lref\schomerus{V. Schomerus, ``D-branes and Deformation Quantization,''
JHEP {\bf 9906}:030 (1999); hep-th/9903205.}
\lref\kapustin{A. Kapustin, ``D-branes in a topologically nontrivial 
B-field,'' hep-th/9909089.} 
\lref\brylinski{J.-L. Brylinski, {\it Loop Spaces, Characteristic 
Classes, and Geometric Quantization}, Prog. in Math. {\bf 107}, 
Birkh\"auser, Boston, 1993.} 
\lref\murray{M.K. Murray, ``Bundle gerbes,'' dg-ga/9407015.}
\lref\careymurray{A. Carey, J. Mickelson, and M.K. Murray, 
``Bundle gerbes applied to quantum field theory,'' hep-th/9711133.}
\lref\doug{R.Douglas, {\it $C^*$-Algebra Extensions and K-Homology,}
Annals of Mathematics Studies, {\bf 95}, Princeton University Press
(1980).}
\lref\berkdoug{M. Berkooz and M. R. Douglas, ``Five-branes in M(atrix)
Theory,'' Phys. Lett. {\bf B395} (1997) 196; hep-th/9610236.}
\lref\bss{T. Banks, N. Seiberg and S. Shenker, ``Branes From Matrices,''
Nucl.Phys. {\bf B490} (1997) 91; hep-th/9612157.}
\lref\chs{C. G. Callan, J. A. Harvey and A.Strominger,
``World-Sheet Approach to Heterotic Instantons and Solitons,''
Nucl.Phys. {\bf B359} (1991) 611; C. G. Callan, J. A. Harvey and
A.Strominger, ``Supersymmetric String Solitons,'' in Trieste
1991, Proceedings, String theory and quantum gravity, hep-th/9112030.}
\lref\seiberg{N. Seiberg, ``A Note on Background Independence in
Noncommutative Gauge Theories, Matrix Model and Tachyon Condensation,''
hep-th/0008013.}
\lref\kuiper{N. H. Kuiper, ``The homotopy type of the unitary group
of Hilbert space,'' Topology {\bf 3} (1965) 19.}
\lref\leigh{D. Berenstein, V. Jejjala, and R.G. Leigh, ``Marginal 
and relevant deformations of N=4 field theories and 
non-commutative moduli spaces of vacua,'' hep-th/0005087}
\lref\vancea{I.V. Vancea, ``On the algebraic K-theory of the 
massive D8 and M9 branes,'' hep-th/9905034}
\lref\matsuo{Y. Matsuo, ``Topological charges of Noncommutative 
soliton,'' hep-th/0009002}
\lref\jmw{D. P. Jatkar, G. Mandal and S. R. Wadia, ``Nielsen-Olesen
Vortices in Noncommutative Abelian Higgs Model,'' hep-th/0007078.}

\Title{\vbox{\baselineskip12pt
\hbox{hep-th/0009030}
\hbox{EFI-2000-28}
\hbox{RUNHETC-2000-33}
}}
{\vbox{\centerline{Noncommutative Tachyons}
\medskip
\centerline{and K-Theory}}}

\smallskip
\centerline{Jeffrey A. Harvey}
\smallskip
\centerline{\it Enrico Fermi Institute and Dept. of Physics}
\centerline{\it University of Chicago}
\centerline{\it 5640 Ellis Ave. Chicago IL, 60637}
\smallskip
\centerline{Gregory Moore}
\smallskip
\centerline{\it Department of Physics, Rutgers University}
\centerline{\it Piscataway, New Jersey, 08855-0849}

\bigskip
We show that the relation between D-branes and
noncommutative tachyons leads very naturally to
the relation between D-branes and K-theory.
We also discuss  some relations between
D-branes and K-homology, provide a noncommutative generalization
of the ABS construction, and give a simple physical interpretation
of Bott periodicity. In addition, a framework for constructing
Neveu-Schwarz fivebranes as noncommutative solitons is proposed.

\medskip

\Date{September 4, 2000}

\newsec{Introduction}

Dbranes can be incorporated into open string field theory
as solitons of tachyon configurations
\refs{\sen,\wittenk,\horava,\hk,\MJMT,\bsz} and carry charges
which take values in K-theory \refs{\mm,\wittenk,\horava}. 
It was recently pointed out
\refs{\dmr,\hklm}
that the description of $D$-branes as
solitons in the open string tachyon field theory
simplifies dramatically when a $B$ field is turned on, thus making
the tachyon field theory into a noncommutative field theory with the
$D$-branes appearing as noncommutative solitons \gms.

One  point of the following paper is that
this description provides another
point of view on the relation between
$D$-branes and K-theory. Indeed this point of
view  makes the relation between $D$-branes and
K-theory manifest.

A second, more speculative point we would like to
make is the following. In the discussion below we will
encounter some simple $C^*$-algebras. It is natural to
wonder if replacing string field algebras by $C^*$-algebras
leads to some new and interesting string backgrounds, or
whether the theory of $C^*$ algebras should play a more
fundamental role in brane physics.

Some   observations closely related to this paper have
been independently made in \refs{\wittenstrings,\periwala}.
Some of the points below were made in lectures at
Strings 2000 \michigantalks. Other recent papers suggesting 
a role of K-homology in D-brane physics include 
\refs{\vancea,\leigh,\periwalb,\matsuo}.

\newsec{Noncommutative tachyons are maps to classifying spaces}

We first consider noncommutative tachyons in the bosonic string.
The basic setup in \hklm\ is that we consider spacetime to
be a product $X \times R^{2n}$, where $X$ is a $26-2n$ manifold.
(\hklm\ take $X$ to be $R^{25-2n,1}$,
but the generalization to arbitrary $X$
is easy, and quite important for our point below.)

We now consider open bosonic string field theory with target
$X \times R^{2n}$.  The action depends on the on-shell
background values of the closed string fields $g_{\mu\nu}, g_s,
B_{\mu\nu}$, where $g_s$ is the closed string coupling.
 We take $g_s$ to be small, and assume that
the natural generalization of the flat space formulae
to curved $g_{\mu\nu}$ applies.

If the tachyon effective action at $B=0$ is:
\eqn\openactp{
S = {C\over g_s} \int_{X\times R^2} d^{26}x \sqrt{\det g} \biggl(
{1 \over 2} f(T) g^{\mu\nu} \partial_\mu T \partial_\nu T -
 V(T) + \cdots \biggr)
}
where $C$ is a constant  and
$T$ is the tachyon field, then the generalization to $B\not=0$
is  given in terms of a noncommutative field theory \refs{\cds,\schomerus,\sw}:
\eqn\openact{
S = {C\over G_s} \int_{X\times R^2}  d^{26}x \sqrt{\det G} \biggl(
\half f(T) G^{\mu\nu} D_\mu T D_\nu T - V(T) + \cdots\biggr)
}
where   $G_s$ and $G_{\mu\nu}$ are the open string
coupling
and metric, given by standard formulae \refs{\ft,\acny,\sw}. 
The effect of $B$ is
to transform $g_s \to G_s, g_{\mu\nu} \to G_{\mu\nu}$ and commutative
products of fields to noncommutative products taken with the Moyal
product. In addition, $B$ induces a non-zero coupling of the tachyon
to the noncommutative $U(1)$ gauge field \garousi.

The tachyon potential is
\eqn\tachpot{
V(T) = V_0 - m^2 T*T + \lambda T*T*T + \cdots
}
There are also higher derivative terms in \openact\ that we have ignored.

The construction  in \refs{\dmr,\hklm} is heavily based on the
noncommutative solitons of \gms.
According to \gms\ the most effective way to think about the tachyon
dependence on the noncommutative directions is in terms of operators on
Hilbert
space. The coordinates $x^{2i-1}, x^{2i}$ on the transverse $R^{2n}$ satisfy
\eqn\noncomm{
[x^{2i-1}, x^{2i}] := x^{2i-1} * x^{2i} - x^{2i} * x^{2i-1} = -i \theta_i
}
where the $\theta^i$ are the skew eigenvalues of the parameter $\theta^{ij}$
appearing in the Moyal product.

Letting $x^a$ denote commutative coordinates along $X$, $x^i$, $i=1,\cdots 2n$
denoting the noncommutative coordinates, the tachyon field $T(x^a,x^i)$
is now regarded as an operator valued function of the $x^a$.
What kind of operator can $T$ be?  Since $T$ is a real field, $T$ should
be a self-adjoint operator.
Since we would like to speak of continuous tachyon fields, $T$ should
be a map of $X$ into a $C^*$ algebra, and since all such
algebras are subalgebras of the algebra of bounded operators on
Hilbert space we regard $T$  as a continuous map:
\eqn\tachmap{
T: X \rightarrow \CB
}
where $\CB$ is the $C^*$-algebra of bounded   operators
on a Hilbert space $\CH$
and we use the norm topology.

In fact, since we wish to have an action,
$T$ should have a derivative\foot{More precisely, $T$ should
have a Frech\'et derivative.}. Moreover, the
gauge fields should be introduced using unbounded operators $D_i = 
\theta^{-1}_{ij} {\rm ad} X_j + {\rm ad} A_i$
on $\CH$. 

After integrating out the massive string fields  the effective action
for the tachyon and gauge fields takes the form
\eqn\openacti{\eqalign{
S = {C\over G_s} \int_X d^{26-2n}x \sqrt{\det G} \biggl(
  {\Tr}\Biggl[ & {1 \over 2}
 f(T)G^{ab} D_a T D_b T  +   \half f(T)
G^{ij}[D_i,T][T,D_j^\dagger] \cr 
& -  V(T) - {1 \over 4} h(T) G^{ik}G^{jl}F_{ij}F_{kl}
- {1 \over 4} h(T) G^{ac}G^{bd}F_{ab} F_{cd} \Biggr] +\cdots \biggr) \cr
}}
Here ${\Tr}$ is the trace of the operator on Hilbert space, $x^a$ run
over
the commuting coordinate directions on $X$.
Evidently, in addition to our other criteria,
certain combinations of the map $T$ in \tachmap\ must
be trace class in order to have a finite action.

Let us now consider the limit of \hklm,
$\ap B_{ij} \to \infty$, or equivalently, $\theta^{ij}/\ap \to 0$,
and consider constant tachyon field configurations $\p_a T=0$.
Then  by rescaling the coordinates to remove $\theta_i$ from
the star product  one sees that the action
reduces to the potential term as $\ap B_{ij} \to
\infty$  and hence $T$ must
satisfy $V'(T) =0$.
 
As noted by \gms\ this can be solved by
\eqn\gmssol{
T= \sum_i \lambda_i P_i
}
where $P_i$ are orthogonal projection operators and $\lambda_i$
are stationary points for $V(T)$.

In the bosonic string formulated in Witten's open string
field theory with $*$ product the potential is purely
cubic. If we assume the basic shape of the potential
remains unchanged after integrating out massive string
fields (recent computations \refs{\ks,\sz,\bsz,\wt} have provided nontrivial
evidence that this is correct), then there are two
stationary points $\lambda=0, \lambda=t_*$.
If we choose $t_*$ to correspond to the perturbative
open bosonic string
vacuum, with $V(t_*)$ given  by the tension of the D25 brane,
then Sen's conjecture states that $V(0)=0$ represents the
closed string vacuum.
Therefore, the only nontrivial constant solution to  \gmssol\
is $T= t_* P_n$ where $P_n$ is a rank $n$ projection operator.

Now, in the limit of \hklm\ the action is proportional to
${\Tr} V(T)= n V(t_*)$ even if the projection operator
$P_n$ varies as we move in $X$. We 
immediately see the close connection to $K$-theory.
Slowly varying tachyonic field configurations are given by
maps from $X$ into the space of rank $n$ projection operators
in Hilbert space. 
This space of projection operators is sometimes denoted
$BU(n)$, so we have 
\eqn\mapbs{
T: X \rightarrow BU(n)
}
If we consider a rank $n<k$ projection operator
in the finite dimensional Hilbert space $C^k$ then the
space of such projection operators is clearly
$U(k)/(U(n) \times U(k-n))$. The space $BU(n)$ is defined
as the inductive limit of this quotient space as $k \rightarrow \infty$.

The space $BU(n)$ is topologically intricate, and if
$X$ is topologically nontrivial then the set of
homotopy classes of maps $[X, BU(n)]$ can be
nontrivial. Indeed, $BU(n)$ is
a model for a ``classifying space'' of vector bundles.
This means there is an isomorphism
\eqn\standiso{
Vect_n(X) \cong [X, BU(n)]
}
where $Vect_n(X)$ are the isomorphism classes of complex vector bundles
on $X$ of rank $n$. This is explained in detail in
\refs{\bottandtu,\atiyahk}. In this way we relate homotopy classes of
tachyon field configurations directly to isomorphism classes of
vector bundles, and therefore to $K$-theory classes.

In the bosonic string the physical interpretation of these
$K$-theory classes is less clear than in type II theory since
the branes carry no conserved charges and presumably are unstable,
even if the $K$-theory class is non-trivial. Our hypothesis is
that these $K$-theory classes label inequivalent unstable
D-brane configurations or boundary states of the bosonic string.

It would be very interesting to extend this discussion
to the case of finite $\theta$ and to include the effects
of second derivatives. Such considerations lead to
many new questions beyond the scope of this paper.
Some of these considerations indicate the relevance
of a nonlinear sigma model with target space $BU(n)$.
\foot{ Such sigma models have been considered in a
superficially different context by Losev, Nekrasov,
and Shatashvili \lns.}

\newsec{Witten's factorization of  the open string $*$ product algebra}

We now consider spacetime of the form $X \times R^2$ with $X$ a
24-manifold. We also assume the metric factorizes and denote
the closed string metric on $X$ by $g_{ab}$ and the closed string
metric on $R^2$ by $g_{ij}$.
Witten has observed
in \wittencomment\  that in the limit  of \hklm,
where the closed string
metric $g_{ij}$ is fixed and  $\ap B_{ij} \to \infty$,
(so the open metric $G^{ij} \to 0$)
the $*$ algebra of open string field theory factorizes as
$\CA \to \CA_0 \otimes \CA_1$. Here
$\CA_0$ is the  algebra of the  vertex
operators in the 26 dimensional
open bosonic string with zero momentum in the
noncommutative directions
and $\CA_1$ is the algebra of noncommutative functions on $R^2$.

We can trivially extend the analysis of \wittencomment\ by considering
the following two scaling limits. In the first we take $B_{ij}=t B^0_{ij}$
and $g_{ab}= t^2 g^0_{ab}$ and take $t \rightarrow \infty$ keeping
$B^0$, $g^0_{ab}$ and $g_{ij}$ fixed. In this limit the string algebra
factorizes as above but with $\CA_0$ the algebra of zero momentum
vertex operators and
$\CA_1=C(X) \otimes C_B(R^2)$ where the first term
is the commutative algebra of functions on $X$ and the second is the
noncommutative algebra of functions on $R^2$ defined by the Moyal
product.
\foot{There is an important question
of whether the functions
should be compactly supported, or not. We believe that
rapid falloff, or compact support is appropriate.}
The second scaling limit takes $B_{ij}=0$ and scales both
$g_{ab}$ and $g_{ij}$ as $t^2$. In this limit the string algebra factorizes
with $\CA_1=C(X \times R^2) = C(X) \otimes C(R^2)$ being the algebra
of commutative functions on $X \times R^2$.

It is natural to expect that the set of D-branes, or
boundary states is somehow connected with a K-theory
of the algebra $\CA_0\otimes \CA_1$. However, since
$\CA_0$ is a vertex operator algebra, the meaning
of its K-theory definitely requires some explanation.
Without answering this question we can at least
ask what we can   say without knowing too much about $K(\CA_0)$.

Our working hypothesis is that
$\CA_0,\CA_1$ behave similar to $C^*$ algebras. In
$C^*$-algebra theory there is    a Kunneth-type
theorem which implies that, modulo torsion, we
may identify $K(\CA_0)\otimes K(\CA_1)$ with
$K(\CA_0 \otimes \CA_1)$. (See
  \blackadar, Theorem 23.1.3.).
Therefore, we will focus on the $K$ theory of the algebra $\CA_1$ in
the next section.

%
%
%

\newsec{Bott periodicity and noncommutative solitons}

The algebra of functions $\CA_1$ is very different for $B=0$
and for $B\not=0$. Nevertheless we expect
the $K$-theory classification of
branes to be unmodified when we turn on $B$ and scale the metric.
We will interpret this statement as a manifestation
of Bott periodicity. (See \segaltalks\ for a related
remark.)

Bott periodicity is usually formulated as
\eqn\bottpi{
K(X) \cong  K(X \times  S^2) = K_{cpt}(X \times R^2)
}
In \hklm\ $X$ is   $R^{23+1}$
 with  $R^2$  as the transverse
2 dimensions to the D23 brane constructed as a
noncommutative soliton of the tachyon field theory.
Equation \bottpi\  can be translated into the algebraic
setting:
\eqn\bottpii{
K(C(X))  \cong  K( C(X)\otimes C_0(R^2) )
}
where $C(X)$ is the algebra of continuous functions
on $X$, and $C_0(R^2)$  is the algebra of continuous
functions going to zero at infinity.

K-theory is
unchanged under ``Morita equivalence.''
Therefore:
\eqn\bottpiii{
K(C(X)) = K(C(X) \otimes Mat_N(C) )
}
Moreover,  the norm-closure of the
$N\rightarrow \infty$ limit of $Mat_N(C)$
is the algebra of compact operators $\CK$.
Since $K$-theory behaves well under
inductive limits,
\eqn\bottpiv{
K(C(X)) = K(C(X) \otimes \CK).
}

If the transverse
coordinates satisfy $[x^1, x^2] = -i \theta$ ($\theta$ is real)
then the Stone-von Neuman theorem says
there is a unique irreducible unitary representation $\CH$,
i.e. the Hilbert space of  quantum mechanics.
Moreover,    to any $f \in \CS(R^2)$,
the Schwarz space of functions of rapid decrease, the
Weyl ordered operators,
\eqn\bottpv{
T(f) = \int dp_1 dp_2 \hat f(p_1,p_2) \exp \bigl[ i (p_1 \hat x^1 + p_2
\hat x^2 )\bigr] ,
}
where $\hat f(p_1,p_2)$ is the Fourier transform, generate
the algebra $\CK$ of compact operators \rieffel.
If we suppose that the classification of $D$-branes is unchanged
in the limit $B\to 0$ then it follows that
$K\bigl(C(X) \otimes C_B(R^2) \bigr) \equiv K(C(X)\otimes \CK) =
K\bigl(C(X)\otimes C_0(R^2)\bigr)$.
Combining this with Morita equivalence we obtain the
statement of Bott periodicity.

\newsec{K-theoretic classification of  D-branes
from tachyons in type IIB Strings}

Let us now turn to the tachyon field in the construction
of type II $D$-branes via noncommutative solitons. We will
focus on the case of BPS IIB branes.
As shown in \wittencomment\ the tachyon field
must satisfy:
\eqn\partisom{
T \bar T T = T
}
where $\bar T$ is the Hermitian conjugate of $T$.
Equation \partisom\ is the defining equation of a ``partial
isometry.'' Moreover, the net brane charge is
given by the index of $T$. In an effective field theory
approach \foot{This result could presumably also be derived
in string field theory} the tachyon potential has the form
\eqn\tveff{V(T,\bar T)=U(\bar T T -1)+U(T \bar T -1) }
To have a finite energy configuration the kernels of both
$T$ and $\bar T$ must be finite dimensional, thus  $T$ should
be both a Fredholm operator and a partial isometry.

Once again, we split spacetime as $X \times R^{2n}_B$,
where $X$ has dimension $10-2n$ and might be topologically nontrivial.
If we consider $X$-dependent configurations with
finite net number of branes then the
tachyon field  will give us  a map
\eqn\partsimi{
T: X \to \CF
}
where $\CF$ are the Fredholm operators. But this is
exactly one model for K-theory!
\atiyahk\janich.
Moreover,  the map $[X, \CF] \to K^0(X) \to 0$ is given by
taking the index bundle whose fiber at $x\in X$ is
just $Ind(T)_x:=Ker(T(x)) - Cok(T(x))$
and we identify this as the $K$-theory class of the
Chan-Paton space of the D-brane. The argument that the map
is onto, given in appendix A of \atiyahk,  shows that
there is no loss of generality in supposing that the
Fredholm operator is in fact a partial isometry.
Thus, one recovers in a very straightforward way
the classification of type II D-brane charge in terms
of K-theory.

A closely related remark has been made (independently)
by Witten in
\wittenstrings\ in the type IIA context. Here Witten uses
the Fredholm model identifying $K^1(X)$ with $[X, \CF^{sa}]$
where $\CF^{sa}$ are the self-adjoint Fredholm
operators. This model is due to Atiyah and
Singer  \asskewadjoint.

\newsec{Toeplitz Operators and the ABS Construction}

In the explicit solution for the D7 brane as a vortex in
the noncommutative plane, explained in
\refs{\hklm,\wittencomment,\jmw} the tachyon operator $T$ is a
special kind of partial isometry, namely, a shift
operator $T=S$ where $S$ is  the shift operator
\eqn\shiftop{
S: |n\rangle  \to  |n+1\rangle   , \qquad  n\geq 0
}
in a ``harmonic oscillator'' basis $\vert n \rangle$, $n\geq 0$, for
a separable Hilbert space.
Note that    $S^*S =1$, but $SS^*$ is not 1, indeed,
$SS^* = 1-\vert 0 \rangle\langle 0 \vert$.
The $C^*$ algebra generated by an operator such
that $S^*S=1$, but $S S^*\not=1$ is unique, and
known   as   the ``Toeplitz algebra.''
This algebra can be realized in several ways,
and the following is particularly apt for
discussing generalizations of noncommutative
tachyons.

We consider our Hilbert space to be
the Hilbert space of square integrable functions on the circle,
$L^2(S^1)$. The functions ${1\over \sqrt{2\pi}} e^{i n \theta}$
define a complete orthonormal basis $|n \rangle$ for $n\in \IZ$.
Given a continuous function $f(\theta)$ we may associate
an operator $M_f: \CH \to \CH$ simply by multiplying
a wavefunction $\psi(\theta)$ by $f(\theta)$. This gives
a representation of the commutative $C^*$ algebra $C(S^1)$
on $\CH$. Now consider the Dirac  operator $D= -i d/d \theta$
and split the Hilbert space into the negative and nonnegative
modes of $D$. Let $P$ be the orthogonal projection onto the positive
subspace $H_+ $ of
$L^2(S^1)$ spanned by $|n \rangle$ with $n \ge 0$. Equivalently, we could
view $P$ as the projection onto the subspace of $L^2(S^1)$
consisting of the boundary values of holomorphic functions.
Then given a function
$f(\theta)$ on $S^1$ we can define a Toeplitz operator which
maps $H_+$ to $H_+$ by
$T_f = PM_f$. Note that if $f$ has negative Fourier modes then
$M_f$ does not preserve $H_+$, and hence the projector $P$ acts
nontrivially. For example, if $f_\ell = e^{i \ell \theta}$,
then $T_{f_\ell}$ is just the shift operator $S^\ell$  for $\ell >0$,
but has a kernel for $\ell<0$.
Quite generally, $(T_f)^\dagger = T_{f^*}$, so $f \to T_f$
preserves the adjoint $*$ action. However, the map
$f$ to $T_f$ is not a homomorphism. Indeed, an easy computation shows
that
\eqn\nothomo{
T_{1} -  T_{f_{\ell}} T_{f_{\ell}^*} = P_\ell
}
is the projection operator onto the first $\ell$ levels in $H_+$.
This is a compact operator, and in general it can be shown that,
while $T_f T_g \ne T_{fg}$, the difference
$T_f T_g - T_{fg}$ is a compact operator.

In what follows, this construction of Toeplitz operators will be
generalized to $L^2$ functions on
odd spheres in order to relate the index  of Toeplitz operators
to the winding  number of ABS configurations.

\subsec{Noncommutative ABS Construction}

Let us now generalize the construction of $T$ in
\refs{\hklm,\wittencomment} allowing
for a $2p$-dimensional transverse noncommutative space\foot{The
construction in \wittencomment\ includes the possibility of a $2p$-
dimenional transverse space for a single $D9$-anti $D9$ pair. Here
we generalize this to $2^p$ pairs in order to explain the relation
between the index of $T$ and the winding number of the ABS configuration.}.

First, we construct the noncommutative tachyon field.
Let us skew-diagonalize $\theta^{ij}$ and take:
\eqn\osc{
[x^{2i-1},x^{2i}] = - i \theta_i \qquad \theta_i > 0 , i = 1, \dots p
}
Moreover, we consider the irreducible Clifford
representation $\gamma_i$ for $\CC \ell_{2p}$.
These are $2^p \times 2^p$ dimensional complex
Hermitian matrices of the form:
\eqn\clifford{
\gamma_i  = \pmatrix{ 0 & \Gamma_i \cr \bar \Gamma_i & 0 \cr}
}
Now we take the noncommutative tachyon to be of the
same form as the commutative ABS configuration \refs{\abs, \wittenk}:
\eqn\spinortach{ T = f(r) \Gamma_i x^i  }
except that we now  regard the tachyon as an operator
\eqn\tach{
T: \CH \otimes S^- \to \CH \otimes S^+
}
where the Hilbert space $\CH$ is realized as
a representation of $p$ oscillators and $S^-, S^+$
are negative and positive spin representations. To be specific,
we will represent $\CH$ as the Bargmann quantization
\eqn\barghil{
\CH_B = Hol(C^p,\exp[-2 \sum \theta_i \vert z_i\vert^2 ] dv)
}
The wavefunctions are holomorphic functions of $z_i= x^{2i-1}+i x^{2i}$,
normalizable with respect to  the above measure and $dv$ is the
standard Euclidean volume element.
An orthogonal basis for $\CH_B$ is provided by the monomials
$z^k:=\prod_i (z_i)^{k_i}$. We will let $k$ stand for a
multiindex $k\in (\IZ_+)^p$.

We now show that $T$ is a Fredholm operator, and also show how
to determine $f(r)$ from the equations $T\bar T T = T$,
$\bar T  T \bar T = \bar T$.
The key calculation is
\eqn\bothways{
\eqalign{
\Gamma_i x^i \bar \Gamma_j x^j & = \sum_{i=1}^p 2\theta_i (N_i + \half)
-i \Sigma_{ij} \theta^{ij} \cr
\bar \Gamma_i x^i  \Gamma_j x^j & = \sum_{i=1}^p 2\theta_i (N_i + \half)
-i \bar\Sigma_{ij} \theta^{ij} \cr}
}
Here $\Sigma_{ij}= {1 \over 4}(\Gamma_i \bar \Gamma_j - \Gamma_j \bar
\Gamma_i)$,
$\bar \Sigma_{ij}={1 \over 4} (\bar \Gamma_i \Gamma_j -
\bar \Gamma_j \Gamma_i )$ and $N_i = a_i^\dagger a_i$ is the $i$th occupation
number.
The second terms in \bothways\ are diagonalized by the spinor
weights to
be $\sum_{i=1}^p \pm \theta_i $. Our convention is that in the second
equation of \bothways\
we have a spinor weight giving $- \sum \theta_i$. Therefore, the first
operator has
no kernel and the second operator has a one dimensional kernel, given by
the
oscillator ground state times the lowest weight spinor.
Thus,
\eqn\teebar{
\bar T = \bar \Gamma_i x^i {1\over \sqrt{ \Gamma_i x^i \bar \Gamma_i x^i
} }
}
satisfies the equation $T \bar T T = T$,
has no kernel and is of index $-1$. We will refer to this as
the ``noncommutative ABS construction.'' In order to explain the
relation to the ABS construction we would like to make sense
of  restricting
the tachyon field to a sphere in the noncommutative space.
Classically, we restrict the field $T$ to the solutions
of the equation
\eqn\sphere{
\sum_i \vert z_i \vert^2 = R^2
}
defining the sphere $\Sigma$ of dimension $2p-1$ and
radius $R$.  At nonzero $B$ field the
$z_i$ become noncommuting, so the question arises as to
what it means to restrict the operator to a noncommutative
sphere.  We will now propose one interpretation of what this
might mean.

In quantum mechanics, restricting the wavefunctions in
the Bargman space $\CH_B$ to the sphere produces the
Hardy space $\CH_\Sigma$. This is the Hilbert subspace of
$L^2(\Sigma;d\Omega)$ defined by the boundary values of
holomorphic functions. Here $d\Omega$ is the standard
round measure on the sphere such that
$dv = R^{2p-1} dR d \Omega$. The projection operator
from $L^2$ to $\CH_\Sigma$ is given by
\eqn\projeop{
\eqalign{
(Pf)(z)  & = \int_\Sigma K_\Sigma(z,w) f(w) d \Omega\cr
K_\Sigma(z,w) & = (1- z\cdot \bar w)^{-p} \cr}
}
An orthogonal basis for the Hardy space is again
given by $\varphi_k = z^k$. Note, that the norm of these
states in the Hardy space is
$$
(z^k, z^{k'}) = \delta_{k,k'} {2 \pi^p \prod_i (k_i)! \over \Gamma(\vert
k \vert + p) }
$$
where $\vert k\vert  = \sum k_i $ for a multi-index $k$.

Now let us consider the action of classical coordinates $z_i$, $\bar
z_i$
on the Hardy space $\CH_\Sigma$. To make sense of this we need to define
Toeplitz operators. In general, if $f: \Sigma \to \IC$ is any function
we define $\CT_f := P M_f$ where $M_f: \CH_\Sigma \to L^2$ is the
operator
of multiplication by $f$. The operators $\CT_{z_i},
\CT_{\bar z_i}$ are easily computed:

\eqn\topelitz{
\eqalign{
\CT_{z_i} \varphi_k & = \varphi_{k+e_i} \cr
\CT_{\bar z_i} \varphi_k &  = 0 \qquad \qquad\qquad\qquad if\quad  k_i =
0 \cr
& = 2\pi {k_i \over \vert k \vert + p-1} \varphi_{k-e_i} \qquad if\quad
k_i > 0 \cr}
}
where $e_i$ is the $i^{th}$ unit vector in $(\IZ_+)^p$.

 By considering the
Hilbert space $\CH_\Sigma \otimes \IC^N$, Toeplitz operators for
functions are easily
generalized
to Toeplitz operators for
matrix valued functions $f: \Sigma \to Mat_N(C)$, and hence we can
consider
our tachyon operator \tach\ above as a Toeplitz operator
\eqn\tachtoep{
T: \CH_\Sigma \otimes S \to \CH_\Sigma \otimes S
}
where $S^+ \cong S^- \cong S$ is the irreducible spin
representation in odd dimensions.
The Toeplitz operator is the projection $P_+$ composed with
matrix multiplication by
$\beta: \Sigma \to GL(N,C)$ given, essentially by the ABS construction:
$$\beta(x) = \Gamma_i x^i {1\over \sqrt{x^i x^i + const.}} $$

The operator $T$ in \tachtoep\ is bounded and  Fredholm.
Now, although the restriction map $\CH_B \to \CH_\Sigma$ is {\it not}
unitary it is 1-1 and onto. Therefore, the {\it index} of $T$
on $\CH_B$ will be the same as the index of $T$ on $\CH_\Sigma$.

Now, we can invoke the index theorem of Boutet de Monvel
\boutet, according to which the index is:
\eqn\bdm{
Index(T\vert_{\CH_\Sigma}) = \int_\Sigma \ch(\beta) Td(T\Sigma)
}
Here
$$
\ch(\beta) = \beta^*(\sum_{j\geq 0} (-1)^{j-1}{\omega_{2j-1}\over (j-1)!} )
$$
and $\omega_i$ are standard generators of $H^i(GL(N,C),Q)$.
Since $Td(T\Sigma) = 1$ in this case we have a direct connection
between the index of the tachyon operator on $\CH_B$,
and the winding number of the classical ABS tachyon.

\newsec{Remarks on the relation to K-homology}

The noncommutative ABS construction in
the previous section leads rather naturally to a relation
between D-branes and the work of Brown, Douglas, and
Filmore (BDF) on the classification of algebras of essentially
normal operators \bdf. In this section we will give a
brief review of that work, and then explain the relation
to D-branes.

\subsec{Brief review of BDF}

Expository discussion of \bdf\ can be found in
\refs{\doug,\blackadar,\baumdouglas,\wegge,\rosenberg}.
For the readers' convenience we give a brief summary
here.

In Matrix theory \bfss, spacetime emerges from an algebra of
commuting operators. Here we will discuss algebras of
``almost commuting'' operators in the belief that these
are related to D-branes.
Recall that by Gelfand's theorem,
$C^*$ algebras of commuting operators are
naturally associated to Hausdorff topological spaces $X$
by considering the algebra of continuous functions $C(X)$.
\foot{If $X$ is noncompact, we add the condition that
$f \to 0 $ at infinity, and correspondingly $C(X)$
does not have a unit. For simplicity of discussion,
we henceforth assume $X$ is compact in this subsection.}
Isomorphism classes of algebras are in 1-1 correspondence
with homeomorphism classes of spaces. We will now
consider noncommutative $C^*$ algebras $\CA$ which
fit into the short exact sequence:
\eqn\exctseq{
0 \rightarrow \CK \rightarrow \CA ~{\buildrel \beta\over \rightarrow} ~ C(X)
\rightarrow 0
}
for some fixed space $X$. Note that if $T_f$ denotes
some operator in $\CA$ mapping to the function $f$
under $\beta$, then
 $T_{f_1} T_{f_2} - T_{f_1 f_2}$ is in the kernel of
$\beta$, and hence is a compact
operator. It follows that $[T_{f_1}, T_{f_2}]$ is
compact and thus the algebra $\CA$ is thus ``almost commuting''
in the sense that compact operators are considered to be ``small.''
An example of
such an extension is given by the Toeplitz  algebra
generated by the shift  operators, described at the beginning
of section 6: $S = T_z \rightarrow z$ defines a $C^*$ morphism
onto the continuous functions on $X=S^1$.

In \bdf\ BDF investigated extensions of the form \exctseq\ for
fixed $X$. To any such extension we can associate a
$C^*$-algebra morphism (called the ``Busby invariant'')
$\tau: C(X) \to Q(\CH)$ where $Q$ is
the ``Calkin algebra'' defined by $Q(\CH) := B(\CH)/\CK$ where
$B(\CH )$ is the algebra of bounded operators on a separable
Hilbert space. Indeed, for any $f\in C(X)$ we choose an
operator $T_f\in \CA$ projecting to it, and define $\tau$ by:
$\tau(f) = \pi(T_f)$ where $\pi: B(\CH) \to Q(\CH)$ is the
projection.  Since $T_{f_1} T_{f_2} - T_{f_1 f_2}$ is a compact
operator, $\tau$ is an algebra homomorphism.  Conversely, given
a $C^*$-algebra morphism $\tau: C(X) \to Q(\CH)$ one can
form an extension \exctseq, and, up to a natural notion of
isomorphism, $\tau$ uniquely characterizes the extension.
Full details can be found in ch. 3 of  \wegge. Suffice it
to say here that, given $\tau: C(X) \to Q(\CH)$ we can form
\eqn\exctseqp{
0 \rightarrow \CK \rightarrow \CA' ~ \rightarrow~ C(X) \rightarrow 0
}
by defining
\eqn\pullback{
\CA' := \{ (\CO, f) : \pi(\CO) = \tau(f) \} \subset B(\CH) \oplus C(X)
}
and that \exctseqp\ is equivalent to \exctseq\ in the sense that
there is an isomorphism $\psi: \CA \to \CA'$ compatible with the
two sequences.

One of the reasons the Busby invariant is useful is that it allows
one to define a notion of direct sum of extensions. In order
to do this we must first introduce ``unitary equivalence,''
also known as ``strong equivalence.''  Two
extensions \exctseq\ are ``strongly equivalent'' if there is
a unitary operator $U$ on $\CH$ such that the
Busby invariants are related by $\tau_2(f) = \pi(U) \tau_1(f) \pi(U)^*$.
Let ${\rm \bf Ext}(C(X),\CK)$ denote the set of strong equivalence
classes of extensions of $C(X)$ by $\CK$.
A direct sum operation on ${\rm \bf  Ext}(C(X),\CK)$ can then be
defined by taking the extension corresponding
to the Busby invariant
\eqn\dirctsum{
\tau_1 \oplus \tau_2: C(X) \rightarrow Q(\CH) \oplus Q(\CH) \rightarrow
Q(\CH \oplus \CH) \cong Q(\CH).
}

It turns out that \dirctsum\ defines a semigroup operation on ${\rm \bf
Ext}(C(X),\CK)$.
 Thus far, the theory
could have been developed for general extensions ${\rm \bf Ext}(A_1, A_2)$
of arbitrary $C^*$ algebras
$A_1$ by $A_2$. However, specializing to $A_1=C(X)$ and $A_2=\CK$, a number
of nice things begin to happen. It turns out
that there is a natural zero in the semigroup, corresponding to the
``trivial extensions.'' These are extensions for which the Busby invariant
lifts to $B(\CH)$; equivalently, they are extensions such that
  the sequence \exctseq\ splits,
and hence we can unambiguously write every operator in $\CA$ in the
form $T_f = N_f + k $ with $k\in \CK$ and $N_{f_1} N_{f_2} = N_{f_1 f_2}$.
Let $Ext(C(X), \CK)$ be the quotient of
${\rm \bf Ext}(C(X),\CK)$ by the trivial extensions.
In the above references it is shown that
 every extension has an inverse (up to the addition of a trivial
extension) so that $Ext( C(X),\CK )$
in fact is an abelian group. Moreover, $Ext( C(X),\CK )$  can even
be used to define a homology
theory on $X$! Indeed, if  we define
\eqn\analytick{
\eqalign{
K_1^a(X) & = Ext( C(X),\CK) \cr
K_0^a(X) & = Ext(C(X) \otimes C_0(\IR),\CK ) \cr}
}
(the superscript ``a'' is for ``analytic,'' K-homology, as opposed to
``topological'' K-homology), then
it turns out that $K_*^a(X)$ is a mod 2 periodic homology theory,
dual to $K$-theory.

One way to make the relation to a homology theory more evident is to
introduce the noncommutative spheres $\CS^0,\CS^1$ with function
spaces
\eqn\noncmsph{
\eqalign{
C(\CS^1) & = \{ \pmatrix{ a_{11} & a_{12} \cr a_{21} & a_{22}\cr} : a_{ij}
\in B(\CH), a_{12}, a_{21}\in \CK \} \cr
C(\CS^0) & = \{ \pmatrix{ a_{11} & 0 \cr 0 & a_{22}\cr} : a_{11}-a_{22}\in
\CK \} \cr}
}
and then define  $K_i(X)$ to be homotopy classes of maps of $X$ into $\CS^i$, 
$K_i(X) := [\CS^i, X] $. In the noncommutative setting this amounts to 
the homotopy classes of $*$-homomorphisms $C(X) \to C(\CS^i)$. (The equivalence 
of this definition to what we described above is hardly obvious. The necessary 
technical details can be found in \blackadar, sections 15.7 and 15.8. )

\subsec{Algebra extensions associated to IIA branes}

We will now review a contruction from
\baumdouglas\ which may be interpreted as
saying that every IIA D-brane naturally provides a nontrivial
extension of the algebra of functions on spacetime
by compact operators.

Let $W$ be an odd-dimensional
$Spin^c$ submanifold of a spacetime
$X$. $W$ is equipped with a complex vector
bundle $E$ with connection and inherits a metric
from $X$. We think of $W$ as the IIA brane
worldvolume and $E$ as its Chan-Paton bundle.
Using the above data  we can form the Hilbert
space $\CH$ of $L^2$ spinors with values in $S\otimes E$,
where $S\to W$ is the spin bundle. Denote the
Dirac operator on $S\otimes E$ by
$\Dsl_E$. Assuming the connection
and metric are generic, $\Dsl_E$ will have no
zeromodes and we can decompose the
Hilbert space into the positive and negative
eigenspaces of $\Dsl_E$:  $\CH = \CH_+\oplus \CH_-$.
The commutative algebra $C(W)$ is represented
on $\CH$ by   multiplication operators
$M_f$ for $f\in C(W)$. In general, $M_f$ does not
preserve the subspace $\CH_+$, but if we take
the ``compression'' of $M_f$ by composing with the
projection operator $P_+:\CH \to \CH_+$ then
we can define a Toeplitz operator
 $T_f= P_+ M_f: \CH_+\to \CH_+$.
As in the  case $W=S^1$
described previously, it turns out that
$T_{f_1} T_{f_2} - T_{f_1 f_2} $
is a compact operator and we obtain an  extension
\eqn\etsn{
0 \rightarrow \CK \rightarrow \CA \rightarrow C(W) \rightarrow 0
}
where $\CA$ is the $C^*$ algebra generated by the $T_f$.
By using pullback  $\phi^*: C(X) \rightarrow C(W)$ we
obtain an extension of the algebra of
functions on all of spacetime.
That is, if $\phi: W\to X$ is a continuous map then
we can define a Busby invariant $\tau \phi^*: C(X) \to Q(\CH)$
from which we   get an extension:
\eqn\etsnp{
0 \rightarrow \CK \rightarrow \widetilde{\CA}  \rightarrow C(X)
\rightarrow 0
}

It is shown in \baumdouglas\ that all classes in
$K_1^a(X) = Ext(C(X),\CK)$ can be obtained from the above
construction using a suitable
triplet $(W,E,\phi)$. Moreover, if a suitable
equivalence relation is put on $(W,E,\phi)$ then
classes in $K_1^a(X)$ are in 1-1 correspondence with
classes $[(W,E,\phi)]$. The equivalence relations
on $(W,E,\phi)$ make good physical sense: they include
cobordism (i.e. continuous deformation of the worldvolume
and Chan-Paton bundle) and a natural identification of
direct sums of Chan-Paton bundles. In addition
they include ``vector bundle modification,''
a mathematical construction reminiscent of the Myers
dielectric effect \myers.

It is interesting to compare $K_1^a(X)$ with
the group of D-brane charges, thought
to be given by $K^1(X)$.
If $X$ is compact, even dimensional  and spin then, modulo
torsion,  $K_1(X)$ is
isomorphic to $K^1(X)$ by Poincar\'e duality. However,
when we include torsion a puzzling difference
emerges. There is a  universal coefficient theorem
(\blackadar, Theorem 16.3.3):
\eqn\univct{
0 ~\rightarrow ~Ext(K^0(X),\IZ) ~ \rightarrow
Ext(C(X),\CK) ~ \rightarrow ~ Hom(K^1(X),\IZ) ~
\rightarrow 0
}
Moreover, the sequence splits, so that the torsion
can in principle differ from that of $K^1(X)$.
This possible discrepancy in torsion charges
deserves to be more thoroughly investigated.

\subsec{The index theorem}

We can now put the noncommutative ABS tachyon field of the
previous section into its proper mathematical context: The
equivalence of IIB D-brane charges in the commutative and
noncommutative theory is simply the equality of the
analytic and topological index, expressed in the framework
of $K$-homology (as explained in \baumdouglas).

In the language of Brown-Douglas-Filmore, the Toeplitz operators
on the  Hardy space
defines an analytic K-homology class
\eqn\khomolgy{
[(\CH_\Sigma, \tau)] \in K_{1,a}(\Sigma^{2p-1})
}
where $\tau$ is the Busby invariant.
That is, the inverse image under $\pi: B(\CH) \to Q(\CH)$
of $\tau(C(\Sigma^{2p-1}))$ in $B(\CH_\Sigma)$
defines an algebra of operators $\CT$ providing a nontrivial
extension by compact operators:
\eqn\exts{
0 \rightarrow \CK \rightarrow \CT \rightarrow C(\Sigma^{2p-1}) \rightarrow
0
}

It is explained in \baumdouglas\ that the K-homology class \khomolgy\
is the same as that determined
by the Dirac operator $[\Dsl]$ on $\Sigma$ using the construction of section
7.2.   In particular,
the index theorem of Boutet de Monvel follows from
the ordinary index theorem.

One usually associates IIB D-brane charge to $K^0(X)$,
or for an infinitely extended D-brane with transverse
space $X_t$, to $K^0_{cpt}(X_t)$. The relation to \khomolgy\
is explained as follows.
We consider the exact sequence in K-homology for
the pair $(D^{2p}, \Sigma^{2p-1})$, where $D^{2p}$ is the
disk of dimension $2p$ with boundary $\Sigma^{2p-1}$.
The connecting homomorphism
gives an isomorphism
\eqn\isom{
\delta: K_0(D^{2p},\Sigma^{2p-1}) \cong K_1(\Sigma^{2p-1})
}
In this way, the above construction associates an element of
analytic $K$-homology $K_{0,a}$ to the noncommutative
tachyon. By Poincare duality   $K_0 \cong K^0$, (again,
up to torsion)
and we produce the same K-theory class we expected to
associate to a IIB brane.

\subsec{Speculations on noncommutative D-branes}

The above considerations lead to the idea that it might
be fruitful to relax the equivalence relations
we have put on the extensions \etsnp. As we have discussed, any
``commutative D-brane'' defines a triple $(W,E,\phi)$
and hence a particular extension. Conversely, given
an abstract extension \etsnp\ could one extract the data of
a D-brane? We can easily answer one simple question
about such generalized D-branes, namely: ``{\it Where} is the
brane?''  as follows.  The
kernel of the Busby invariant $\tau: C(X) \to Q$
defines an ideal, and from the Gelfand corespondence
therefore defines a subspace $W\subset X$. Concretely,
the ideal is the subalgebra of functions vanishing on
$W$. It would be natural to identify $W$
with the worldvolume of a D-brane.
Whether or not one can usefully recover other
aspects of the structure of a D-brane, and in particular
whether extensions \etsnp\ which do {\it not} come from
triples $(W,E,\phi)$ can be usefully identified with
``noncommutative D-branes'' remains an interesting
open question.

In any case,
inspired by the result of BDF we would like to define an
action whose solutions could be considered to be the
set of possible IIA D-branes, generalized in the above sense.
The action has some interesting similarities to the
IKKT action.
On the other hand, we caution the reader at the outset
that it remains to be seen  if the following action will play
any useful role in the computation of any physical quantities.

The action is a function
of pairs $(\CA, \phi)$, where $\CA$ is a $C^*$ algebra
and $\phi $ is a $C^*$-algebra morphism
$\phi: \CA \to C(X)\to 0$, and is given by
\eqn\action{
S(A,\phi) := {\rm sup}_{f_1, f_2\in C(X)} {\rm inf}_{\phi(a_i)= f_i}
{\Tr}_D
([a_1, a_2] [a_1, a_2]^\dagger).
}
Here we first take the infimum over all lifts $T_f$ of a pair of
functions $f$. Moreover,  ${\Tr}_D$ is the Dixmier trace.
Roughly speaking, ${\Tr}_D$ is defined as follows.
Let $\mu_n(T)$ be the eigenvalues of $\sqrt{T^\dagger T}$
arranged in decreasing order. Define
\eqn\dixmier{
{\Tr}_D(T) := \lim_{N\to \infty} {1\over \log N} \sum_{n=1}^N \mu_n(T).
}
For the real story, consult the book by Connes \connes.

The action \action\ is positive semidefinite. So, in
any reasonable ``space of $(\CA, \phi)$'' the
zeroes of the action are automatically stationary points
of minimal action. The action   is identically zero
only when, for all $f_1, f_2\in C(X)$ there are lifts
$T_{f_1}, T_{f_2}$ such that
the commutator $[T_{f_1}, T_{f_2}]$ has singular
values falling off faster than $1/\sqrt{n}$.
We may expect the relations \topelitz\  to give a good
approximation to the general behavior of
$[T_{f_1}, T_{f_2}]$ on spinors of large energy,
and from this we expect that
the extensions associated to $(W,E,\phi)$ described
above will be zeroes of the action.
 Conversely, any zero of the action can be used to
define an extension of $C(X)$ by
compact operators.

It is interesting to compare the action \action\ with the
IKKT model:
\eqn\ikkt{
S = g_{IK} g_{JL} {\rm Tr}\biggl( [X^I, X^J] [X^K, X^L]\biggr)
}
where $X^I$ are $N\times N$ Hermitian matrices and $g_{IJ}$ is a
nondegenerate constant metric on $\IR^{10}$.
If we consider the $X^I$ as generators of the algebra of
functions on $\IR^9$ then there is a certain similarity
between \action\ and \ikkt.
However we note that

1. The IKKT action does not generalize easily to curved
spaces. Even on $\IR^9$ if we attempt to include curved
metrics $g_{IJ}$ we run into ordering problems.
(See \douglasnc, for the state of the art on this problem.)

2. When producing D-branes from Matrix theory the solutions
have infinite action. Of course, this is physically appropriate
for infinitely extended planar branes. Nevertheless, it would
be nice to work with finite action quantities when considering
compact branes.

\newsec{Nonzero $H$-fields and 5-branes}

In this section we will focus on a description of Neveu-Schwarz
fivebranes in the framework of \hklm. We should first discuss what we mean
by an NS fivebrane in open string theory. In the original description
\chs\ NS fivebranes are solutions to the closed string equations of
motion with topology $M \times S^3 \times R$ with $M$ the fivebrane
world volume such that  $\int_{S^3} H = Q_5$, $H$ being the NS three-form
field and $Q_5$ the quantized fivebrane charge. Since the tension
scales like $1/g_s^2$ with $g_s$ the closed string coupling, these
are properly thought of as solitons in the closed string sector rather
than the open string sector of the theory where soliton energies scale
as $1/g_s$ (as for D-branes). In open string theory we cannot expect
to see the detailed form of the closed string solution since closed
string states only arise at the loop level in open string theory.
We thus define a fivebrane to be a configuration in a ten-dimensional
spacetime $X$ with $H \in H^3(X,Z)$ a non-trivial integer class.

Given the scaling of the fivebrane tension with $g_s$, the close connection
between the framework of \hklm\ and Matrix theory \seiberg, and the well-known
difficulties in describing fivebranes in Matrix theory \refs{\berkdoug,\bss},
we can anticipate some problems here as well. 

To explain the basic idea and the difficulty one expects, consider taking
$X=W  \times R^2_B$ to be the world-volume of an unstable $D9$-brane
in IIA with a large B-field on the $R^2$ component and take $W=R^5
\times S^3$. $W$ represents the commutative part of the
$D9$-brane world volume.  The effective 
action  \openacti\ contains gauge fields with gauge group $U({\cal H})$
coupled to the tachyon field in the adjoint representation. Since the
$U(1)$ component of $U({\cal H})$ (with $A_\mu$ proportional to the
identity operator) does not couple to $T$ and has infinite action if
its field strength is non-zero, it is more correct to say that the
gauge group is $PU({\cal H}) \equiv U({\cal H})/U(1)$. Defining
more precisely what is meant by the $U(1)$ component when there
is non-trivial topology is quite subtle as will be discussed below. 

Since $U({\cal H})$ is contractible \kuiper, 
$\pi_2(PU({\cal H}))= \pi_1(U(1))=\IZ$. 
Thus we can construct an ``instanton'' configuration of the
$PU({\cal H})$ gauge fields on $S^3$ by patching together gauge fields
on the northern and southern hemispheres using a non-trivial element
of $\pi_2(PU({\cal H}))$ on the $S^2$ equator. Our proposal is that such
a twisted $PU({\cal H})$ bundle with the tachyon field $T=t_*$ represents
a $D9$-brane in the presence of a NS fivebrane while condensing the
tachyon field to $T=0$ removes the $D9$-brane and leaves an NS fivebrane.
More generally, we can use non-trivial projection operators for
the tachyon to study lower $D$-branes in the presence of NS fivebranes.

We can now see one difficulty we expect to encounter. Since the NS
fivebrane world volume is six-dimensional, it must span $R^5 \in W$ and as well
have one component in the noncommutative plane 
$R^2_B$. As a result, the trace in \openacti\
is expected to diverge, i.e. the 
gauge field fieldstrength-squared for  the twisted
$PU({\cal H})$ connection is not expected to be trace class. More precisely,
we expect that if we cut off the trace by summing over a finite number
of modes then the trace will diverge   in the mode number cutoff. 
In fact, an evaluation of the gauge action $\int_W {\Tr} F\wedge *F$
for  some examples of smooth nontrivial 
$PU(\CH)$ connections shows that the action is in indeed infinite. 
A proper interpretation of this infinity must be addressed in 
future work. Here we simply note that since the mode-number 
cutoff can be interpreted as an infrared cutoff in the 
noncommutative directions, there is room for an interpretation 
of the infinite gauge kinetic action as the volume divergence 
due to the extension of the 5-brane worldvolume in the noncommutative 
directions.

To see the connection to the fivebrane definition in terms of $H$, we
note that a standard theorem states that principal
$PU({\cal H})$ bundles are
classified by the Dixmier-Douady class
$h\in H^3(W,Z)$. 
\foot{ A quick homotopy-theoretic proof is that  
$BPU(\CH)   \sim K(\IZ,3)$ since $\Omega B PU(\CH) \sim 
PU(\CH) \sim K(\IZ,2) \sim \Omega K(\IZ,3)$. }
This class has been interpreted in
\refs{\bouwknegt,\atiyahsegal,\wittenstrings} 
as the cohomology class of the $H$-field
of string theory. In addition to the arguments 
presented in these papers we would like to point 
out that the reasoning described by Kapustin in 
\kapustin\ for the case when $h$ is torsion in 
fact can be extended to the case of $h$ non-torsion. 
This follows because a nontrivial $PU(\CH)$ bundle 
defines a nontrivial ``bundle gerbe with connection,'' 
(where we are using the terminology explained in 
\murray\careymurray). Then, using the equivalence to 
the formulation of Brylinski \brylinski\ one 
can argue that the ``holonomy of the $PU(\CH)$ connection 
in the fundamental representation'' can be given a 
concrete definition in terms of a covariantly 
constant section of a line bundle with connection 
over loop space $L W$. The line bundle with connection 
over $LW$ is constructed using the bundle gerbe 
associated to the $PU(\CH)$ bundle with connection $A$
in a way explained in \brylinski\murray\careymurray. 

In more physical terms, we wish to make sense of 
the expression 
\eqn\anomcanc{
\exp\biggl[ i \int_D B \biggr] {\Tr}_{\CH} P \exp \int_\gamma A
}
in the open string path integral, where $D$ is the disk worldsheet 
with boundary $\gamma \subset W$,  $B$ is the background 
$B$-field, and $A$ is a $PU(\CH)$ connection. In order to 
define this we must lift $A$ to a compatible $U(\CH)$ 
connection $\tilde A$. In so doing the field strength acquires 
a ``$U(1)$ component'' which we denote by 
${\Tr} \tilde F$, although since we are working with 
  operators not necessarily of trace class this notation 
should be handled with great care. The essential physical point 
is that in infinite dimensions the commutator of two 
Hermitian operators can be proportional to the identity 
matrix, the standard example being a pair of operators 
representing  the Heisenberg relations. Consequently   the 
commutator term in $\tilde F = d \tilde A + \tilde A^2$ 
can in fact contribute to the $U(1)$ component of $\tilde F$ 
and in topologically interesting situations it must do so. 
This in turn means that  the  Bianchi identity for the $U(1)$ part of 
$\tilde F$ is {\it not} $d {\Tr} \tilde F=0$ but rather 
$d {\Tr} \tilde F = K$ where $K$ is a globally 
well-defined 3-form on $W$. Moreover, by the 
general results of \brylinski\ it follows that 
the 
cohomology class of $K/(2\pi i)$ coincides 
with the Dixmier-Douady class $h$. Defining the 
holonomy of $A$ as a covariantly constant section of 
a bundle over loopspace one can follow the strategy 
of \kapustin\ and conclude that  the Dixmier-Douady
 class $h$ must be identified with that of 
the physical $H$-field. It is not necessary to 
assume that $h$ is a torsion class, although in 
infinite dimensions the trace ${\Tr}$ isolating 
the $U(1)$ part of the field strength requires 
an ad-hoc definition \brylinski. 
 
One simple example of a nontrivial $PU(\CH)$ connection 
illustrating some of the above general remarks is the following. 
(This example is a paraphrase of section 4.3 of \brylinski.) 
We will take the base space to be the three-manifold 
$S^2 \times S^1$, more appropriate to an $H$-monopole. 
A similar (but more elaborate) example applies directly 
to $S^3$ and can be extracted from   \careymurray. 

We will construct a $PU(\CH)$ bundle over $S^2 \times S^1$ 
by starting with a $U(1)\times \IZ$ bundle over 
$S^2 \times S^1$ and then embedding the $U(1)\times \IZ$ 
transition functions into $PU(\CH)$. The $U(1)\times \IZ$
bundle over $S^2 \times S^1$ will simply be  $S^3 \times \IR$  
with a rightaction by $U(1)\times \IZ$ given by:  
\eqn\identfs{
\eqalign{
(u, x) & \sim (u e^{i \chi \sigma^3/2} , x) \cr
(u,x) & \sim (u, x+1)\cr}
}
Here $u\in S^3$ is identified with an $SU(2)$ matrix, the first 
line is the right $U(1)$ action and the second line is the $\IZ$ action 
on $x\in \IR$.  Note that the $S^3$ is {\it not} to be thought of 
as embedded in spacetime. Rather, $W= \IR^5 \times S^2\times S^1$. 

We now consider the Heisenberg algebra generated by 
operators 
$\hat \theta$, and $\hat N$ acting on functions 
in $L^2(S^1)$. This $S^1$ should be thought of as the 
fiber in the Hopf fibration $S^3 \to S^2$. Let 
$\hat \theta$ be the position operator
and $\hat N$   the integrally-quantized angular momentum, 
so that $[\hat \theta, \hat N]= i $. 
Using these operators we can form a representation of 
$U(1) \times \IZ$ in $PU(\CH)$ via: 
\eqn\represent{
(e^{i \chi}, n) \rightarrow e^{i n \hat \theta} e^{i \chi \hat N} 
}
Note that 
$e^{i n \hat \theta}$ and $ e^{i \chi \hat N} $ commute 
up to the phase $e^{i n \chi}$ and hence \represent\ is 
indeed a representation of the commutative group 
$U(1)\times \IZ$ in $PU(\CH)$. Using \represent\ we convert 
the transition functions of the $U(1) \times \IZ$ bundle 
$S^3 \times \IR \to S^2 \times S^1$ into $PU(\CH)$ transition 
functions. 
Of course, we can (by construction) 
lift the transition functions to $U(\CH)$ over contractible 
open sets in a good cover of $S^2 \times S^1$, but then they will
fail to satisfy the cocycle condition. 

We now discuss how to isolate the $u(1)$ part. Technically, this 
is defined in \brylinski\ by the splitting of an exact 
sequence of bundles of the adjoint representation. Here 
the relevant 
Lie algebra of operators is generated by the invariant 
elements $(\hat N - x {\bf 1} )$
and ${\bf 1}$. Note that $\IZ$, being discrete, has no Lie 
algebra. Therefore, we do not need to include $\hat \theta$.
Note too that we are forced to choose the combination 
$(\hat N - x {\bf 1} )$ so that $x\sim x+1$ is equivalent to 
conjugation by $e^{i \hat \theta}$. 
We define ``the $u(1)$ part'' to be the coefficient of 
${\bf 1}$ in this basis.

As an example of a nontrivial $PU(\CH)$ connection we choose 
standard Euler angle coordinates $(\phi, \theta, \psi)$ for  $u\in S^3$ and 
$x$ on $\IR$. Then we may  define the  connection using the 
globally defined Lie algebra valued form  on $S^3 \times \IR$ given by:  
\eqn\connection{
\eqalign{
A_+(u,x) & 
= i \bigl( d \psi_+ + \half (1- \cos\theta)d \phi\bigr)(\hat N - x {\bf 1} ) \cr
A_ -(u,x) & 
= i\bigl( d \psi_- - \half (1+ \cos\theta)d \phi\bigr)(\hat N - x {\bf 1} ) \cr}
}
where we have divided $S^3$ into two hemispheres labelled by $\pm$. 
One may easily check that $A(u, x+1) = e^{ i \hat \theta} A(u, x) e^{-i \hat \theta}$
so this defines   a connection on a bundle over $S^2 \times S^1$. 
According to our definition of the $U(1)$ part of $F$ we have 
 ${\Tr}(F) =   - i A dx $. This is globally defined on $S^3 \times S^1$ but 
is not basic. It is also     not closed, as 
promised, but $K = d {\Tr}(F)\wedge dx  = -{i\over 2}  \sin\theta d \theta d\phi dx$ 
is  a  basic  form, giving the globally defined ``gerbe curvature 3-form'' 
on $S^2 \times S^1$ with $\int_{S^2\times S^1} K/(2\pi i) =1$.

In view of the above, we believe that by allowing for twisted
$PU(\CH)$ bundles in the formalism of \hklm\ we are 
able to include the effects of NS 5-branes in the picture of
\hklm.   Indeed, if $ {\bf \rm A}  \to W$ is
a twisted bundle
with fiber given by $\CK$ and Dixmier-Douady
class $h$ then $\Gamma({\bf \rm A})$ is an algebra
whose (Grothendieck group of) finitely generated projective modules
define $K_H(W)$. In the limit of large
noncommutativity the tachyon field still
defines a projection operator, hence a projective module
for this algebra. In the context of type II strings it is 
important to note that since $PU(\CH)$ also acts on Fredholm 
operators, there is also a Fredholm model for $K_H(W)$.

Obviously, many details need to be worked out in 
the above proposal. We hope to report on this elsewhere.

\bigskip
\centerline{\bf Acknowledgments}

We would like to thank E. Diaconescu, E. Martinec and E. Witten for
discussions and comments. We thank I.M. Singer for useful remarks on 
the manuscript.  We are especially grateful
to G.B. Segal for many discussions and for some prophetic
lectures at the ITP in August 1999 \segaltalks.
We also recommend lectures by I.M. Singer \singertalks.

We would
like to thank the Aspen Center for Physics for hospitality
during the completion of this paper. The work of J.H. is
supported in part by NSF grant PHY-9901194.
The work of G.M. is supported by DOE grant DE-FG02-96ER40949.
\bigskip

\listrefs

\bye